\documentclass[fleqn,10pt]{wlscirep}
\usepackage{array}
\usepackage{makecell}
\usepackage{tabularx}
\usepackage{booktabs}
\usepackage[utf8]{inputenc}
\usepackage[T1]{fontenc}
\usepackage{algorithmic}
\usepackage{braket}
\usepackage{algorithm}
\usepackage{amsmath}
\usepackage{graphicx}
\usepackage{subcaption}
\usepackage{caption}
\usepackage{ulem}
\usepackage{multirow}
\usepackage{siunitx}  
\usepackage{booktabs} 

\usepackage{makecell}
\title{Enhanced Distributed Variational Quantum Eigensolver for Large-Scale MaxCut Problem}

\author[1,+]{Yuefeng Lin}
\author[2,+]{Kun Wang}
\author[3,4,+]{Qinyuan Zheng}
\author[3,4]{Rui Zhang}
\author[3]{Jing-Kai Fang}
\author[1]{Tiejun Meng}
\author[1]{Jingen Xiang}
\author[1,*]{Cong Guo}
\author[3,*]{Jun-Han Huang}

\affil[1]{Shenzhen SpinQ Technology Co., Ltd, Shenzhen, 518048, China}
\affil[2]{College of Physics and Electronic Engineering, Center for Computational Sciences, Sichuan Normal University, Chengdu 610066, China}
\affil[3]{State Key Laboratory of Genome and Multi-omics Technologies, BGI Research, Shenzhen 518083, China}
\affil[4]{School of Artificial Intelligence, University of Chinese Academy of Sciences, Beijing 101408, China}

\affil[*]{Correspondence: cguo@spinq.cn (C.C.), huangjunhan@genomics.cn (J.-H.H.)}

\affil[+]{These authors contributed equally to this work.}

\begin{abstract}
MaxCut is a canonical NP-hard combinatorial optimization problem in graph theory with broad applications ranging from physics to bioinformatics. Although variational quantum algorithms offer promising new approaches that may eventually outperform classical schemes, they suffer from resource constraints and trainability issues such as barren plateaus, making large-scale instances intractable on noisy intermediate-scale quantum devices. 
In this paper, we propose an enhanced distributed variational quantum eigensolver for large-scale MaxCut problems, which extends our prior distributed variational quantum eigensolver framework by integrating a novel hybrid classical-quantum perturbation strategy, enhances optimization scalability and efficiency.  Our algorithm solves weighted MaxCut instances with up to 1000 vertices using only 10 qubits, and numerical results indicate that it consistently outperforms the Goemans-Williamson algorithm. 
We further employ a warm-start initialization strategy, seeding the algorithm with high-quality solutions from the Goemans-Williamson algorithm, with results confirming that the optimal classical solution can be effectively further improved.
The practical utility of the proposed algorithm is further validated through its application to haplotype phasing on genome sequencing data of the human \textit{ABCA1} gene, producing high-quality haplotypes that rival those obtained by the Goemans-Williamson algorithm with $10^6$ projections.
These results establish the proposed algorithm as a scalable, NISQ-compatible framework for near-term quantum-enhanced large-scale combinatorial optimization.

\end{abstract}

\begin{document}

\flushbottom
\maketitle
%
%
\thispagestyle{empty}


\section{Introduction}

    The MaxCut problem, a cornerstone of combinatorial optimization, seeks to partition the vertices of an undirected graph into two disjoint subsets that maximize the number of edges between them. As a canonical NP-hard problem, MaxCut not only occupies a central place in complexity theory but also underpins numerous real-world applications, including VLSI circuit layout, spin-glass modeling in statistical physics, correlation clustering, portfolio optimization, image segmentation, and haplotype phasing \cite{lengauer2012combinatorial, hagen1992new, barahona1982computational, lucas2014ising, shi2000normalized, von2007tutorial, lippert2002algorithmic}. To tackle this challenge, a suite of classical algorithms has been developed, ranging from exact methods that are limited to small instances to powerful heuristics like the Goemans-Williamson (GW) algorithm\cite{goemans1995improved} which deliver a guaranteed 0.878-approximation in polynomial time via semidefinite programming and randomized rounding. It remains NP-hard to approximate MaxCut within a factor of $16/17$\cite{haastad2001some}. 

    For large-scale datasets, MaxCut becomes computationally intractable for classical heuristics, despite their effectiveness on smaller instances. This fundamental barrier drives the search for new computational paradigms, particularly quantum computing, which holds promise for outperforming classical methods on such hard problems. 
    In the current noisy intermediate-scale quantum (NISQ) era\cite{preskill2018quantum}, variational quantum algorithms (VQAs) \cite{mcclean2016theory, cerezo2021variational, tilly2022variational, bharti2022noisy}  represent the leading hybrid quantum-classical framework for tackling combinatorial optimization tasks. By employing a parameterized quantum circuit, also named ansatz, whose parameters are iteratively optimized by a classical optimizer to minimize or maximize a problem-encoded cost function, VQAs strike an attractive balance between expressiveness and hardware compatibility. Such as the variational quantum eigensolver (VQE)\cite{VQE}, originally developed for finding molecular ground states but readily adapted to Ising-type objectives, and the quantum approximate optimization algorithm (QAOA) \cite{farhi2014quantum}, specifically designed for discrete combinatorial problems such as MaxCut. Both approaches are relatively resilient to noise when operated at shallow circuit depths, making them among the most viable candidates for near-term quantum advantage. 
    
    VQAs in the NISQ era face two primary challenges in the form of leveraging limited quantum resources to tackle industrially relevant problems and mitigating the barren plateau phenomenon. To address the resource constraints, distributed quantum computing has emerged as a promising solution \cite{du2022distributed,situ2024distributed, dunjko2018computational,matsuo2020problem,li2022large,xie2025efficient}. The approach relies on a divide-and-conquer strategy, interconnecting and coordinating multiple small-scale quantum processing units (QPUs) to combine their qubit resources for solving larger-scale problems. For instance, our prior work \cite{PRXLife.2.023006} introduced a distributed quantum variational eigensolver for binary optimization problems, which leverages Hamiltonian and trial wave function decomposition to solve an $(n \times p)$-qubit genomic assembly problem using only $n$ qubits. 
    
    However, such distributed frameworks are also plagued by the barren plateaus, a pervasive challenge in variational quantum algorithms, where the cost function's gradient vanishes exponentially as the number of qubits increases ~\cite{mcclean2018barren, wang2021noise, cerezo2021cost,fontana2023adjoint}.
    This decay severely impedes the classical optimization loop, often trapping it in suboptimal local minima and limiting scalability. The issues of vanishing gradients and local optima are likewise prevalent in classical optimization. In metaheuristic algorithms, a common and effective countermeasure is the deliberate introduction of perturbations. For instance, simulated annealing employs temperature-controlled stochastic acceptance ~\cite{kirkpatrick1983optimization}, genetic algorithms utilize mutation operators ~\cite{holland1992adaptation}, and variable neighborhood search relies on systematic neighborhood shifts~\cite{hansen2001variable,mladenovic1997variable}. These strategies have all proven effective on challenging problems such as MaxCut and the traveling salesman problem (TSP)~\cite{cordeau2001tabu,rego2001perturbation}.  In variational quantum algorithms, an analogous and highly effective perturbation-like technique is warm-starting. It leverages a high-quality classical solution to prepare a biased initial state, thereby accelerating convergence, facilitating escape from local minima, and increasing robustness against barren plateaus. Numerical experiments on small-scale graph instances further confirm that the warm-start method consistently outperforms the standard Goemans-Williamson algorithm\cite{egger2021warm, tate2023warm}. 
   
    In this work, we propose an enhanced distributed variational quantum eigensolver (EDVQE) for large-scale MaxCut problems. The algorithm extends our prior DVQE framework by integrating a novel hybrid classical-quantum perturbation strategy, which enhances optimization scalability and efficiency. 
    To assess the efficacy of the proposed EDVQE, we performed comprehensive numerical experiments on dense MaxCut instances, including weighted complete graphs and weighted cluster graphs, with graph sizes spanning 100 to 1000 nodes, while leveraging merely 10 qubits of quantum hardware. Furthermore, we explored a warm-start initialization approach, wherein EDVQE was seeded with approximate solutions obtained from classical solvers and subsequently optimized via the same perturbation method. Finally, to further validate its practical utility, we successfully apply the proposed EDVQE algorithm to the haplotype phasing, a challenging combinatorial optimization task in bioinformatics.
    
    The main contributions of this work are presented as follows:
    
    \begin{enumerate}[label=(\roman*), itemsep=0pt, parsep=0pt]
        \item  A hybrid classical-quantum perturbation mesethod rooted in the MaxCut problem is proposed and integrated with the DVQE algorithm, enabling the solution of large-scale instances using merely 10 qubits, which  significantly reduces hardware requirements while improving algorithmic performance.
        
        \item The simulation results demonstrated that the hybrid classical-quantum perturbation method markedly enhances the performance of the DVQE algorithm. Consequently, the EDVQE algorithm exhibits superior performance in terms of solution quality and approximation ratio compared to classical benchmarks across various graph types and scales.

        \item Simulation results reveal that the warm-start methodology consistently outperforms the Goemans-Williamson algorithm.
        
        \item The implementation of a variational quantum algorithm yields notable results on the gene orientation problem within computational biology, thereby underscoring the real-world applicability of EDVQE.
    \end{enumerate}

\section{Preliminaries}

\subsection{The MaxCut Problem}

Given an undirected graph $G = (V,E)$ with $|V| = N$ vertices and weights \( w_{i,j} \) (where \( w_{i,j} = w_{j,i} \) for all \( (i,j) \in E \)), the objective of MaxCut is to partition the vertex set \( V \) into two disjoint subsets \( S \) and \( \overline{S} \) such that the total weight of the edges crossing between \( S \) and \( \overline{S} \) (i.e., the cut size) is maximized.

To mathematically model the MaxCut problem, we introduce a binary variable \( x_i \in \{-1, +1\} \) for each vertex \( i \in V \), where \( x_i = -1 \) signifies that vertex \( i \) belongs to subset \( S \), and \( x_i = 1 \) indicates assignment to \( \overline{S} \). The problem can then be formulated as a quadratic unconstrained binary optimization (QUBO) objective, aiming to maximize the cut size expressed as:

\begin{equation}
C(\mathbf{x}) = \sum_{(i,j) \in E} w_{i,j} (x_i + x_j - 2x_i x_j),
\label{eq:QUBO}
\end{equation}
where \( \mathbf{x} = \{x_1, \dots, x_N\} \in \{-1,1\}^N \).

A prevalent strategy for solving binary optimization problems  on quantum hardware involves encoding the objective function into a problem Hamiltonian within the Ising model framework. This is accomplished by mapping each binary variable \( x_i \) to a linear combination of the Pauli-Z operator and the identity matrix:

\begin{equation}
x_i \rightarrow \frac{I - Z_i}{2},
\label{eq:binary_to_matrix}
\end{equation}
where \( Z_i \) denotes the Pauli-Z operator acting on the \( i \)-th qubit, and the corresponding spin variable \( z_i \in \{-1, +1\} \) represents the eigenvalue of \( Z_i \). Substituting this mapping into the binary optimization problems expression yields the equivalent Ising Hamiltonian:

\begin{equation}
H_C = \frac{1}{2} \sum_{(i,j) \in E} w_{i,j} (I - Z_i Z_j),
\label{eq:HC}
\end{equation}
and its ground state corresponds to the solution of the original MaxCut problem.

\subsection{Haplotype phasing}
Haplotype phasing aims to reconstruct the allele sequences at heterozygous sites such as single nucleotide polymorphisms (SNPs) on the two homologous chromosomes from sequenced fragments or reads of a single individual, namely the two haplotypes (H1 and H2) derived from maternal and paternal origins. As diploid organisms, humans possess two homologous copies of each chromosome, which exhibit different alleles at heterozygous sites. Fragments obtained from conventional sequencing are randomly derived from the two chromosomes and intermixed, thus making direct distinction of their origins impossible. By leveraging overlaps between fragments and allele consistency, fragments can be assigned to the two haplotypes, thereby physically separating the variant phases on maternal and paternal chromosomes. 

The analysis of haplotypes has proven particularly valuable in elucidating the genetic basis of many diseases. For complex diseases such as diabetes, cardiovascular diseases, and cancer, which involve interactions between multiple genetic and environmental factors, haplotype-based association studies offer greater power to detect susceptibility loci compared to approaches focusing on SNPs. By identifying haplotypes that are significantly associated with disease risk, researchers can gain crucial insights into the underlying genetic architecture and pathogenesis. This knowledge, in turn, provides a foundation for advancing strategies in early diagnosis, risk prediction, and targeted therapeutic interventions.

Traditionally, haplotype information has been inferred statistically from population genotype data using linkage disequilibrium, as implemented in algorithms like PHASE\cite{stephens2001new}. However, the switch error rate is relatively high (approximately 5.4\%) in unrelated individuals\cite{marchini2006comparison}. With the advancement of high-throughput sequencing technologies, particularly the emergence of long fragments and paired-end sequencing, direct haplotype phasing from sequencing data of a single individual has become feasible. The HapCUT algorithm \cite{bansal2008hapcut} proposed by Bansal and Bafna (2008) represents an efficient and accurate combinatorial optimization method that transforms the problem into iteratively solving MaxCut on associated graphs, minimizing the minimum error correction (MEC) score. On the HuRef personal genome data, it achieved assembly accuracy significantly superior to greedy heuristics (20-25\% lower MEC scores) and previous methods, reducing the switch error rate to 1.1-1.4\%. This advancement provides more reliable complete haplotype information for personalized genomics, disease association studies, and precision medicine.

\section{Method}

In this section, we outline the methodology of the proposed algorithm, beginning with the framework of an EDVQE for MaxCut problem.  A concise overview of the DVQE algorithm is then provided. Concluding this section, we present a detailed description of the proposed hybrid classical-quantum perturbation strategy.

\subsection{Framework of Enhanced Distributed Variational Quantum Eigensolver}
 The proposed framework employs an iterative optimization strategy that alternates between classical and quantum computations, systematically perturbing and refining the solution to enhance global exploration and convergence. The overall workflow is depicted in Figure~\ref{fig:framework} and comprises the following key stages:
\begin{figure}[!ht]
    \centering
\includegraphics[width=1.0\linewidth]{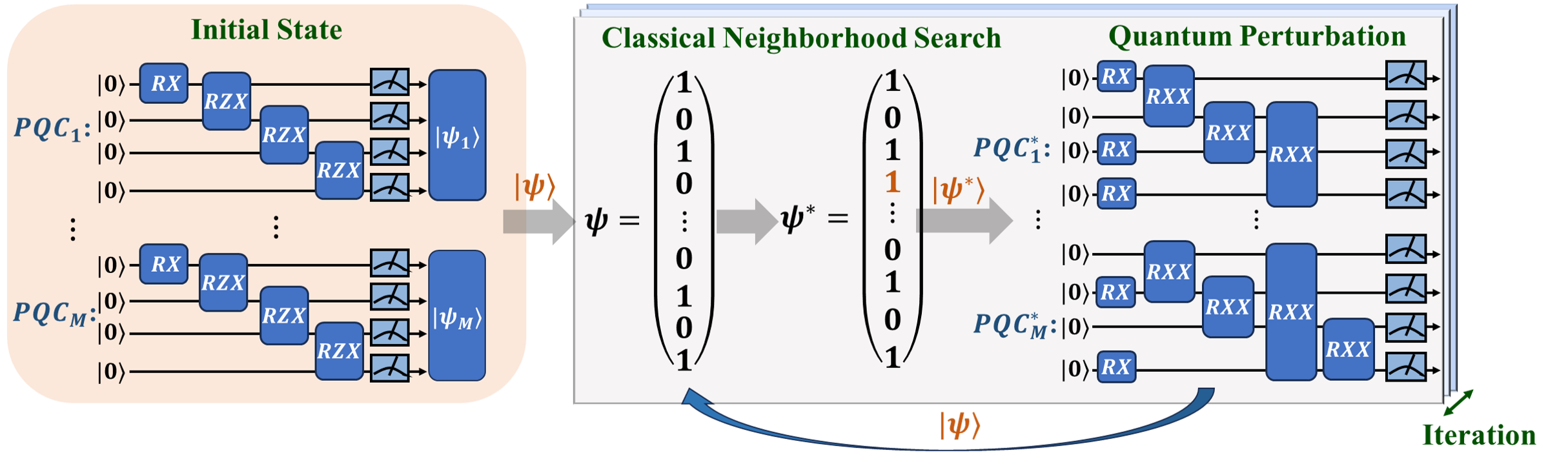}
    \caption{The framework of EDVQE algorithm for MaxCut Problem.}
    \label{fig:framework}
\end{figure}
\begin{enumerate}
    \item \textbf{Initial Solution Generation}: Execute DVQE to compute an initial solution $\ket{\psi}$ by preparing a parameterized quantum state and optimizing its variational parameters to approximate the ground state of the problem Hamiltonian.
    
    \item \textbf{Classical Neighborhood Search}: Starting from $\ket{\psi}$, apply a classical neighborhood search algorithm  (e.g., one based on single-vertex flips) to perform a local exploration, yielding an improved solution $\ket{\psi^*}$.
    
    \item \textbf{Quantum Perturbation}:
    \begin{itemize}
        \item \textbf{Solution Encoding}: Encode the classically refined solution $\ket{\psi^*}$ as the biased initial state in the quantum circuit. This is achieved by applying parameterized $R_X(\phi)$ gates to the corresponding qubits, where $\phi$ is initially set to $\pi$ for qubits that need to be flipped from $\ket{0}$ to $\ket{1}$ to map the binary assignments of $\ket{\psi^*}$ into the quantum state.
        \item \textbf{Quantum Two-Vertex Swap Perturbation}: Introduce parameterized $R_{XX}$ entanglement gates between qubit pairs that correspond to vertices in different partitions (i.e., one in $\ket{0}$ and the other in $\ket{1}$). These gates facilitate probabilistic swaps of partition assignments, enabling exploration of more diverse solution spaces.
        \item \textbf{Parameter Optimization}: Employ a classical optimizer to jointly tune the parameters of the $R_X$ and $R_{XX}$ gates, directing the perturbation toward lower-energy states and potentially escaping local optima.
    \end{itemize}
    
    \item \textbf{Iterative Refinement}: Feed the perturbed solution back into the classical neighborhood search and repeat Steps 2 and 3 until the solution quality stabilizes over consecutive iterations, indicating convergence.
\end{enumerate}
This iterative hybrid approach leverages the strengths of both classical heuristics for efficient local refinement and quantum perturbations for global search, making it particularly suitable for large-scale MaxCut instances on NISQ devices.

\subsection{Distributed Variational Quantum Eigensolver}
The variational quantum eigensolver is a notable quantum-classical hybrid algorithm that leverages the potential of quantum computation in the NISQ era. In the VQE approach, a quantum parameterized circuit, known as an ansatz, is used to prepare a trial quantum state that approximates the ground state of a specified Hamiltonian.
This ansatz is iteratively refined through a classical optimization algorithm, which updates its parameters to minimize the energy expectation value, effectively solving for the ground state.  Based on the variational principle of quantum mechanics, it guarantees that for any parameterized trial state $\ket{\psi(\theta)}$, the expectation value of the Hamiltonian $H$ satisfies
\begin{equation}
    \braket{H} = \bra{\psi(\theta)} H \ket{\psi(\theta)} \geq E_0,
\end{equation}
where $E_0$ denotes the true ground-state energy corresponding to the eigenstate $\ket{\psi_{0}}$.

VQE has shown considerable promise in applications such as quantum chemistry and combinatorial optimization, particularly for simulating molecular systems and solving constrained optimization problems on near-term quantum hardware. Nevertheless, resource requirements for tackling practical scale problems often exceed the capabilities of current NISQ devices. For example, simulating the ground state of a MaxCut Hamiltonian defined on a graph with $|V| = N $ vertices generally requires a quantum circuit with $N$ qubits. To address this problem, 
our prior work has demonstrated the practical feasibility of this modular approach for solving binary optimization problems, such as MaxCut.

In practice, the Hamiltonian 
$H$ for binary optimization problems is typically expressed as a linear combination of terms:
\begin{equation}
    H = \sum_{t} c_t H_t,
\end{equation}
where $c_t \in \mathbb{R}$ and each term $H_t$ takes the form of a tensor product of Pauli operators:
\begin{equation}
    H_t = \bigotimes_{i=1}^M \tilde{O}_{t,i}, \quad \tilde{O}_{t,i} \in \{I, Z\},
\end{equation}
where $\tilde{O}_{t,i}$ denotes the operator acting on the $i$-th subsystem.

Since all Hamiltonian terms $H_t$ are diagonal in the computational basis, they are mutually commuting and thus simultaneously measurable. For any tensor product state $\ket{\psi} = \otimes_{i=1}^M \ket{\psi_i}$, the expectation value of each Hamiltonian term factorizes exactly as:
\begin{equation}
    \bra{\psi} H_t \ket{\psi} = \prod_{i=1}^M \bra{\psi_i} \tilde{O}_{t,i} \ket{\psi_i}.
\end{equation}

 Moreover, each state of the subsystem $\ket{\psi_1}, \ket{\psi_2}, \dots, \ket{\psi_i}, \dots, \ket{\psi_M}$ is prepared and optimized on
 smaller quantum processors, while the global energy expectation is computed through the factorized expression.
\begin{equation}
\begin{aligned}
    \min_{\vec{\theta}} E(\vec{\theta}) &= \min_{\vec{\theta}} \bra{\psi(\vec{\theta})} H \ket{\psi(\vec{\theta})} \\
    &= \min_{\vec{\theta}} \sum_{t} c_t \prod_{i=1}^M \bra{\psi_i(\vec{\theta_i})} \tilde{O}_{t,i} \ket{\psi_i(\vec{\theta_i})}.
\end{aligned}
\end{equation}


Our approach not only preserves the exactness of expectation value calculations despite the distributed architecture, but also significantly enhances scalability by partitioning large binary optimization instances into manageable subproblems. The commutative property of the Hamiltonian terms further simplifies the measurement process, making DVQE a particularly resource-efficient framework for tackling large-scale combinatorial optimization problems on emerging quantum hardware.

\subsection{Hybrid Classical-Quantum Perturbation}

This subsection introduces the hybrid classical-quantum perturbation strategy, a core component of our proposed algorithm that synergistically combines a classical neighborhood search for local intensification with a quantum perturbation for global diversification.

\begin{enumerate}
    \item \textbf{Classical Single-Vertex Neighborhood Search (denoted as CNS-1)}. The classical component of our perturbation strategy is grounded in the well-established framework of Neighborhood Search, which operates on the principle of iterative refinement through local exploration.
    Starting from an initial solution encoded in \(N\) bits, it systematically generates a candidate set by flipping the state of each single vertex while preserving all others. This procedure defines a local neighborhood of cardinality \(N+1\), from which the optimal candidate is selected based on the objective function evaluation. 
    \item \textbf{Quantum Two-Vertex Swap Perturbation (denoted as QP-2)}. Extending the perturbation to two-vertex operations classically would require evaluating the complete \(N^2\) neighborhood, involving \(\frac{N(N-1)}{2} + 1\) distinct solutions. This quadratic scaling of computational complexity \(O(N^2)\) becomes a fundamental bottleneck for large-scale problems. 
    To overcome this limitation while enabling powerful global exploration, we devise the QP-2 strategy, which employs parameterized quantum gates to implicitly explore a vast neighborhood. Specifically, we introduce parameterized \(R_{XX}(\theta)\) gates between qubit pairs that span different graph partitions (i.e., a pair with one qubit in \(\ket{0}\) and the other in \(\ket{1}\)). The matrix representation of the \(R_{XX}(\theta)\) gate is given by:
\begin{equation}
RXX(\theta) = \exp\left(-i\frac{\theta}{2}X \otimes X\right) = 
\begin{bmatrix}
\cos(\frac{\theta}{2}) & 0 & 0 & -i\sin(\frac{\theta}{2})\\
0 & \cos(\frac{\theta}{2}) & -i\sin(\frac{\theta}{2}) & 0\\
0 & -i\sin(\frac{\theta}{2}) & \cos(\frac{\theta}{2}) & 0\\
-i\sin(\frac{\theta}{2}) & 0 & 0 & \cos(\frac{\theta}{2})
\end{bmatrix},
\end{equation}

where $\theta = 0$ yields the identity operation $I \otimes I$, and $\theta = \pi$ results in $-iX \otimes X$, performing a SWAP operation up to a global phase.
The parameter \(\theta\) for each gate is initialized from a uniform distribution \(U[-0.01\pi, 0.01\pi]\) to ensure the system starts from a state that is only slightly perturbed from the CNS-1 solution. The subsequent variational optimization of these \(\theta\) parameters enables probabilistic swapping of partition assignments, thereby facilitating exploration of more diverse solution spaces.
\end{enumerate}

\section{Result}
\subsection{Experimental Setup}
To systematically evaluate the performance of the proposed algorithms, we conduct numerical experiments on two types of weighted graphs, including weighted complete graphs and customized cluster graphs. For each graph type, we generate instances with node counts ranging from 100 to 1000 in increments of 100, resulting in a total of 20 distinct MaxCut instances. All graphs are constructed using a fixed random seed of 3587 to ensure reproducibility.
In the EDVQE algorithm, a parameterized quantum circuit constructed from $R_X$ and $R_{ZX}$ gates is employed to prepare the trial quantum state, as illustrated in the left panel of Figure~\ref{fig:framework}. The parameters of the parameterized quantum circuit are iteratively updated via a classical optimization algorithm until convergence, yielding an initial solution $\ket{\psi}$.
To ensure statistical robustness in our experiments, each graph instance was evaluated over 10 independent runs with randomly initialized parameters, thereby generating varied initial states $\ket{\psi}$ and correspondingly diverse initial solutions.
The performance is assessed using the normalized average cut value, defined as:
\begin{equation}
    \bar{A} = \frac{Cut_1 + Cut_2 + \cdots + Cut_{10}}{10\times|E|} = \frac{1}{10} \sum_{i=1}^{10} \frac{Cut_i}{|E|},
\end{equation}
where \(\mathrm{Cut}_i\) denotes the cut value obtained from the \(i\)-th parameter initialization and \(|E|\) represents the total number of edges in the graph. 
Given the computational infeasibility of obtaining exact solutions for large-scale instances, the classical Goemans-Williamson algorithm was employed as a benchmark, with each graph instance computed using 10 independent runs to obtain the normalized average cut value $\bar{A}$ for comparative analysis.
All simulations are executed using the SpinQit\cite{SpinQit} framework with the Torch simulator, while the optimization processes are performed using the Adam optimizer. In the EDVQE implementation, each subsystem was configured with \(n = 10\) qubits.

\subsection{Performance on Weighted Complete Graphs}

We first evaluate the algorithms on weighted complete graphs, in which every pair of distinct vertices is connected by an edge with weights sampled uniformly from the interval between 1 and 10.

\begin{figure}[!ht]
    \centering
    \includegraphics[width=0.8\linewidth]{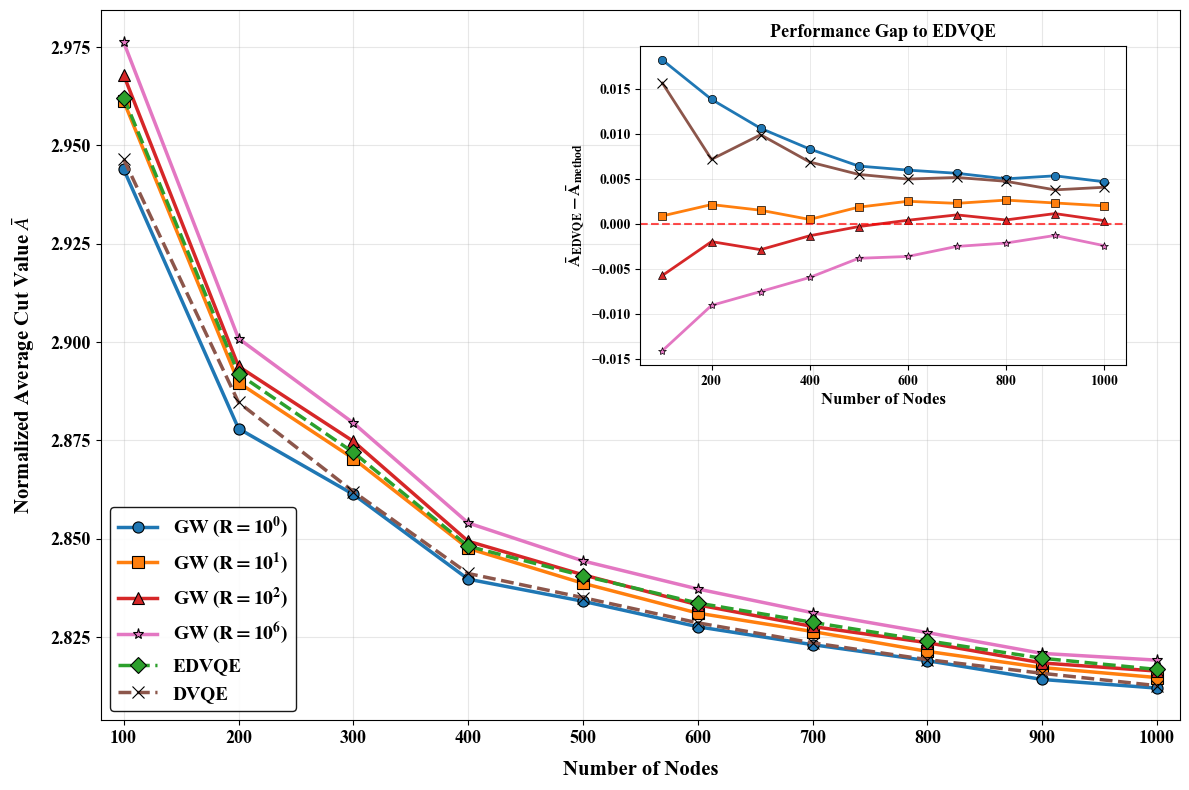}
    \caption{Comparison of EDVQE and GW algorithm with different projections under different nodes in weighted complete graphs.}
    \label{fig:complete_performance}
\end{figure}

The performance of both algorithms, in terms of the normalized average cut value $\bar{A}$, is summarized in Figure \ref{fig:complete_performance} across a range of vertex counts. The results clearly demonstrate the superiority of the EDVQE algorithm over the standard GW algorithm under various projection configurations. Specifically, for problem sizes ranging from 100 to 1000 vertices, the EDVQE without hybrid classical-quantum perturbation, which represents the initial solution, consistently outperforms the standard GW algorithm employing a single projection. This observation demonstrates the inherent advantage of the base DVQE framework.  Moreover, integrating hybrid classical-quantum perturbation within the EDVQE framework yields supplementary performance enhancements. To establish a comprehensive benchmark, we enhance the GW algorithm by increasing its number of projections, which correspondingly elevated its solution quality. As a result, our EDVQE approach maintains consistent superiority over the GW variant with 10 projections and, notably, outperforms the GW algorithm with 100 projections across instances ranging from 600 to 1000 vertices.

To further analyze the performance gains within the EDVQE framework, we break down the contributions of its three key components, which include the initial solution, the classical single-vertex neighborhood search (CNS-1), and the quantum two-vertex perturbation (QP-2). Table \ref{table: Weighted_Completed_Graphs} summarizes the normalized average cut values $\bar{A}$ of the initial solution, along with the absolute improvements from CNS-1 ($\Delta$CNS-1) and QP-2 ($\Delta$QP-2), together with their corresponding percentage enhancements. 
The results indicate that CNS-1 provides a consistent yet modest absolute improvement, with an average $\Delta\text{CNS-1}$ of 0.0025 (a relative gain of 0.089\%). In contrast, QP-2 contributes a more substantial enhancement, delivering an average $\Delta\text{QP-2}$ of 0.0043 (a relative gain of 0.150\%)—approximately double that of the classical component. These findings demonstrate that both post-processing modules consistently improve upon the initial solution, with the quantum-based perturbation yielding notably greater gains.

\begin{table}[!ht]
\centering
\caption{The performance gain within the EDVQE algorithm under different nodes in weighted complete graphs. Here $\Delta$CNS-1 = $\bar{A}_{\text{CNS-1}} - \bar{A}_{\text{Initial}}$ represents the improvement from classical single-vertex neighborhood search, while $\Delta$QP-2 = $\bar{A}_{\text{QP-2}} - \bar{A}_{\text{CNS-1}}$ quantifies the additional enhancement from quantum perturbation.}
\setlength{\tabcolsep}{12pt}
\begin{tabular}{lrrrrrrr}
\toprule
\multirow{2}{*}{\# of Nodes} & \multicolumn{4}{c}{Value} & \multicolumn{2}{c}{\% Improvement} \\
\cmidrule(lr){2-5} \cmidrule(lr){6-7} 
& GW $(R=1)$ & Initial & $\Delta$CNS-1 & $\Delta$QP-2 & CNS-1 & QP-2 \\
\midrule
100  & 2.944 & 2.946 & 0.0056 & 0.0100 & 0.190 & 0.339 \\
200  & 2.878 & 2.885 & 0.0021 & 0.0051 & 0.073 & 0.177 \\
300  & 2.861 & 2.862 & 0.0028 & 0.0068 & 0.098 & 0.238 \\
400  & 2.840 & 2.841 & 0.0032 & 0.0045 & 0.113 & 0.158 \\
500  & 2.834 & 2.836 & 0.0017 & 0.0038 & 0.060 & 0.134 \\
600  & 2.827 & 2.829 & 0.0025 & 0.0026 & 0.088 & 0.092 \\
700  & 2.823 & 2.824 & 0.0019 & 0.0029 & 0.067 & 0.103 \\
800  & 2.818 & 2.819 & 0.0020 & 0.0028 & 0.071 & 0.099 \\
900  & 2.814 & 2.816 & 0.0017 & 0.0022 & 0.060 & 0.078 \\
1000 & 2.812 & 2.813 & 0.0019 & 0.0023 & 0.068 & 0.082 \\
\midrule
Average & 2.845 & 2.847 & 0.0025 & 0.0043 & 0.089 & 0.150 \\
\bottomrule
\end{tabular}
\label{table: Weighted_Completed_Graphs}
\end{table}

\subsection{Performance on Weighted Cluster Graphs}
The tensor product structure adopted in the EDVQE framework results in a global trial state that is devoid of entanglement across different subsystems. Entanglement is instead within each individual subsystem, implemented by designing quantum circuits that incorporate two-qubit gates for every subsystem.

\begin{figure}[!ht]
    \centering
\includegraphics[width=0.8\linewidth]{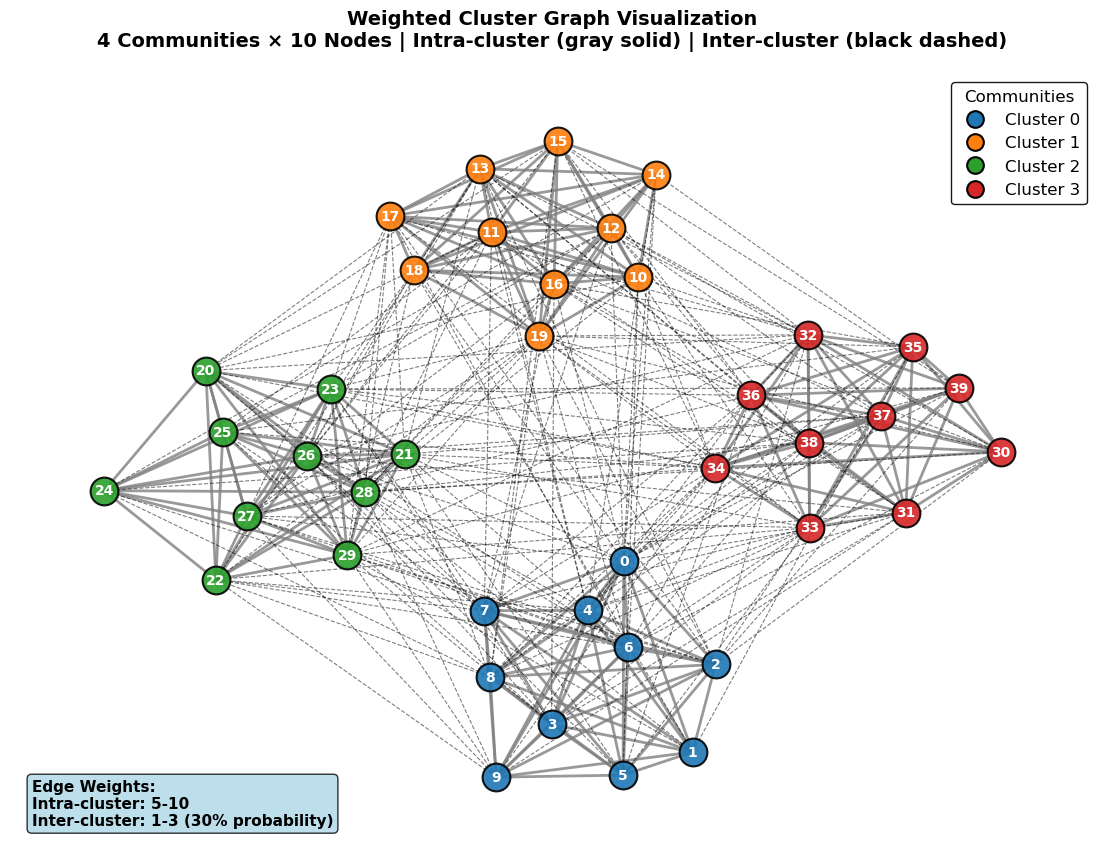}
    \caption{Visualization of the weighted cluster graph benchmark with modular structure, comprising 4 distinct communities of 10 nodes each.}
\label{fig:Cluster_Graphs_4_Communities}
\end{figure}

To evaluate the performance of the inherent entanglement scheme in our algorithm, we further conduct additional simulations on weighted cluster graphs, which serve as a natural benchmark for optimization problems with inherent community structure.
As illustrated in Figure \ref{fig:Cluster_Graphs_4_Communities}, the benchmark weighted cluster graphs are generated by forming distinct clusters with dense intra-cluster connections and sparse inter-cluster links. In detail, intra-cluster edge weights are sampled uniformly from the interval between 5 and 10, while inter-cluster edges are assigned lower weights are sampled uniformly from the interval between 1 and 3 with a 30\% connection probability between different communities. 

\begin{figure}[!ht]
    \centering
\includegraphics[width=0.8\linewidth]{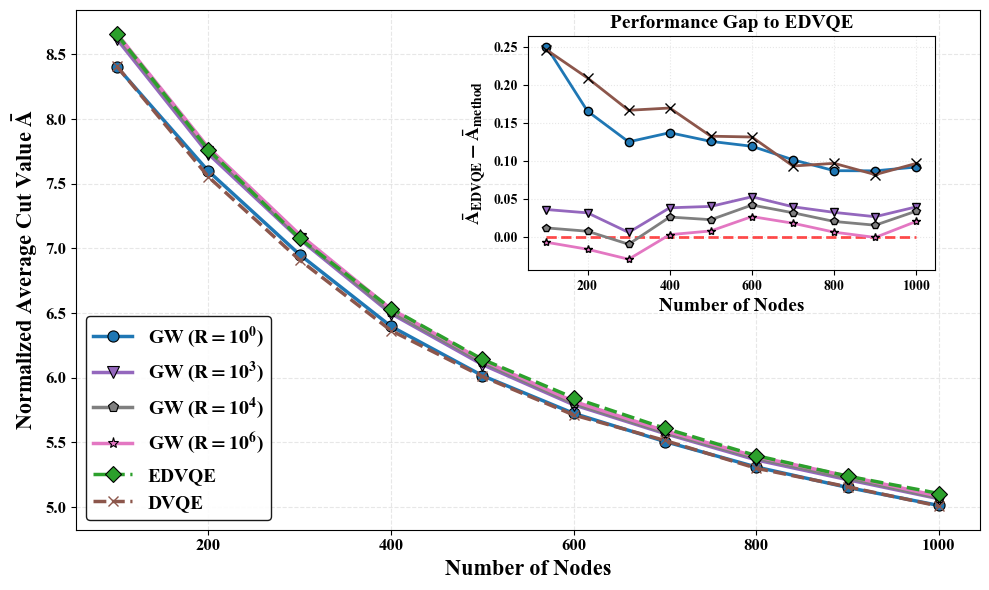}
    \caption{Comparison of EDVQE and GW algorithm with different projections under different nodes in weighted cluster graphs.}
\label{fig:Cluster_Graphs}
\end{figure}

The EDVQE demonstrates more pronounced performance advantages on weighted cluster graphs, as illustrated in Figure~\ref{fig:Cluster_Graphs}. While the initial solution alone is outperformed by the standard GW algorithm, the subsequent application of hybrid classical-quantum perturbation yields a substantial performance improvement. As a result, across nearly all graph sizes (except for instances with 300 nodes), the EDVQE consistently surpasses the GW algorithm using $10^4$ projections. Even more remarkably, for larger-scale graphs ranging from 400 to 1000 nodes, it achieves results that surpass those of the GW algorithm utilizing an extensive $10^6$ projections, underscoring its scalability and robustness in handling structured optimization challenges with community features.

Table \ref{Weighted_Cluster_Graphs} presents a component-wise analysis of the performance improvements in the EDVQE. The CNS-1 yields minimal improvement, with an absolute gain $\Delta\text{CNS-1}$ of 0.0050, corresponding to a relative gain of 0.085\%. In contrast, the QP-2 contributes a substantially larger enhancement, attaining an absolute $\Delta\text{QP-2}$ of 0.1386 and a relative gain of 2.168\% relatively, exceeding the classical component by more than 25 times. This pronounced performance gap underscores the quantum module's distinctive capability to identify cross-community optimal configurations via coordinated multi-qubit operations, thereby effectively addressing the limitations of classical local search in modular graph topologies.

\begin{table}[!ht]
\centering
\caption{The performance gain within the EDVQE algorithm under different nodes in weighted cluster graphs.}
\setlength{\tabcolsep}{12pt}
\begin{tabular}{lrrrrrr}
\toprule
\multirow{2}{*}{\# of Nodes} & \multicolumn{4}{c}{Value} & \multicolumn{2}{c}{\% Improvement} \\
\cmidrule(lr){2-5} \cmidrule(lr){6-7}
& GW(R=1) & Initial & $\Delta$CNS-1 & $\Delta$QP-2 & CNS-1 & QP-2 \\
\midrule
100  & 8.402 & 8.407 & 0.0044 & 0.2423 & 0.052 & 2.882 \\
200  & 7.598 & 7.554 & 0.0035 & 0.2062 & 0.046 & 2.730 \\
300  & 6.952 & 6.911 & 0.0053 & 0.1619 & 0.077 & 2.342 \\
400  & 6.398 & 6.365 & 0.0050 & 0.1652 & 0.079 & 2.596 \\
500  & 6.017 & 6.010 & 0.0037 & 0.1295 & 0.062 & 2.155 \\
600  & 5.725 & 5.713 & 0.0055 & 0.1265 & 0.096 & 2.215 \\
700  & 5.507 & 5.515 & 0.0047 & 0.0892 & 0.085 & 1.617 \\
800  & 5.337 & 5.300 & 0.0058 & 0.0901 & 0.109 & 1.700 \\
900  & 5.152 & 5.158 & 0.0058 & 0.0841 & 0.112 & 1.630 \\
1000 & 5.012 & 5.007 & 0.0065 & 0.0908 & 0.130 & 1.814 \\
\midrule
Average & 6.210 & 6.194 & 0.0050 & 0.1386 & 0.085 & 2.168 \\
\bottomrule
\end{tabular}
\label{Weighted_Cluster_Graphs}
\end{table}

\subsection{Performance of Warm-Start EDVQE}
To further investigate the efficacy of warm-starting, we utilized the best solutions obtained from 10 independent executions of the GW algorithm with $10^6$ projections as the initial configuration for EDVQE, followed by refinement via the classical-quantum hybrid perturbation mechanism.
This methodology is applied to three distinct categories of MaxCut instances, including in the previously examined weighted complete graphs and weighted cluster graphs, we incorporate weighted 3-regular graphs, where edge weights for the latter were uniformly sampled from the interval between 1 and 10.

\begin{figure}[!ht]
    \centering
    \includegraphics[width=1.0\linewidth]{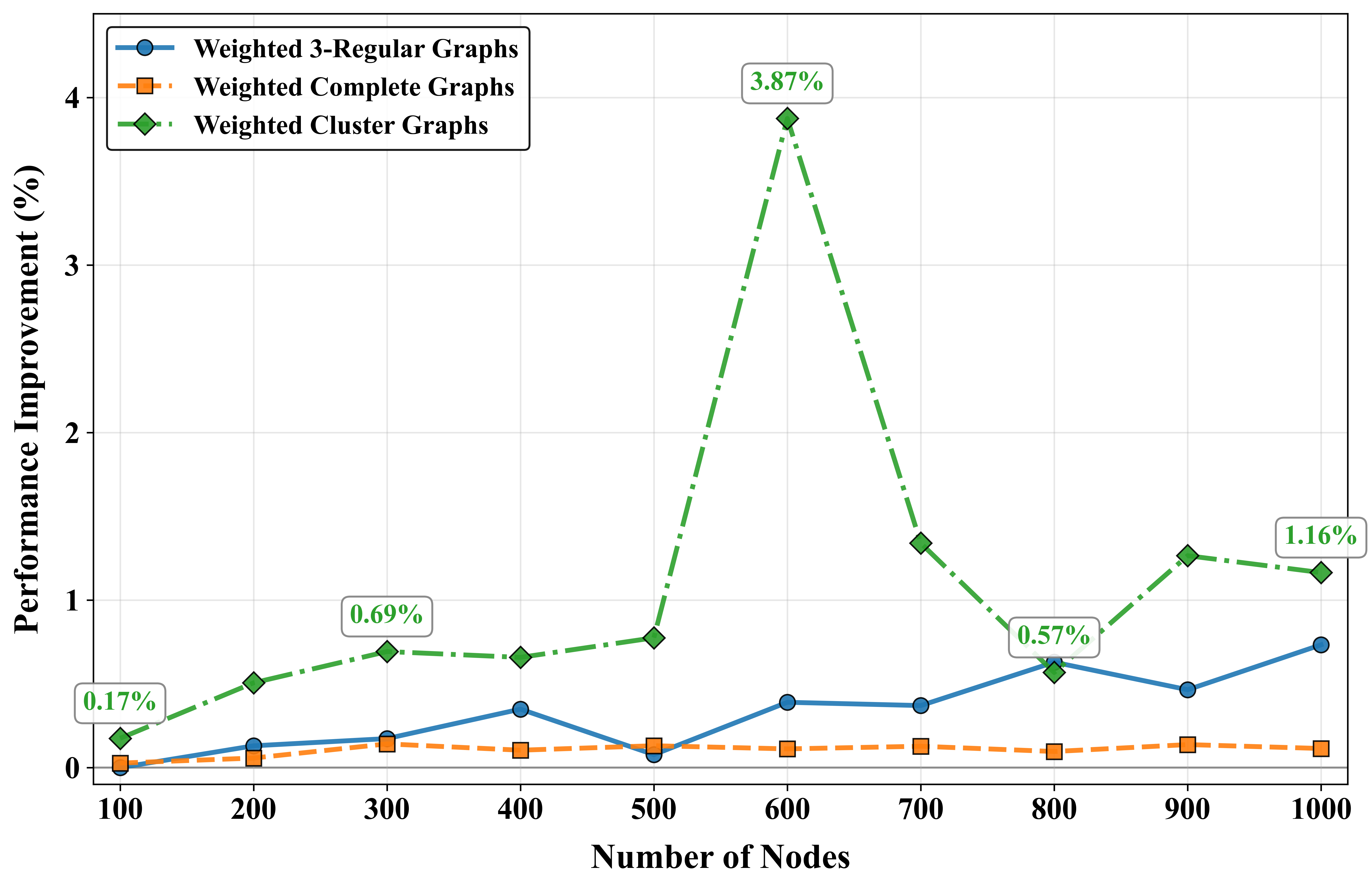}
    \caption{Performance improvement of EDVQE over the GW algorithm across three graph types.}
    \label{fig:edvqe_performance_improvement_fixed}
\end{figure}

Figure \ref{fig:edvqe_performance_improvement_fixed} demonstrates the consistent superiority of the optimized solutions over the GW with $10^6$ projections baseline across all problem scales, with particularly pronounced enhancements observed on dense graph structures. Among the three graph types investigated, cluster graphs exhibit the most substantial performance improvement, achieving an average gain of 1.102\% with peak improvements reaching 3.87\% at 600 nodes. Complete graphs show moderate yet consistent enhancements, maintaining an average improvement of 0.150\% across all tested scales. In contrast, 3-regular graphs display relatively limited optimization potential, with an average improvement of only 0.088\%. This performance hierarchy suggests that the effectiveness of our approach is strongly influenced by graph topology, with more complex and interconnected structures benefiting more significantly from the hybrid optimization framework.

\subsection{Performance for Haplotype phasing}

\begin{figure}[!ht]
\centering
\captionsetup{justification=justified,singlelinecheck=false}
\includegraphics[width=1.0\linewidth]{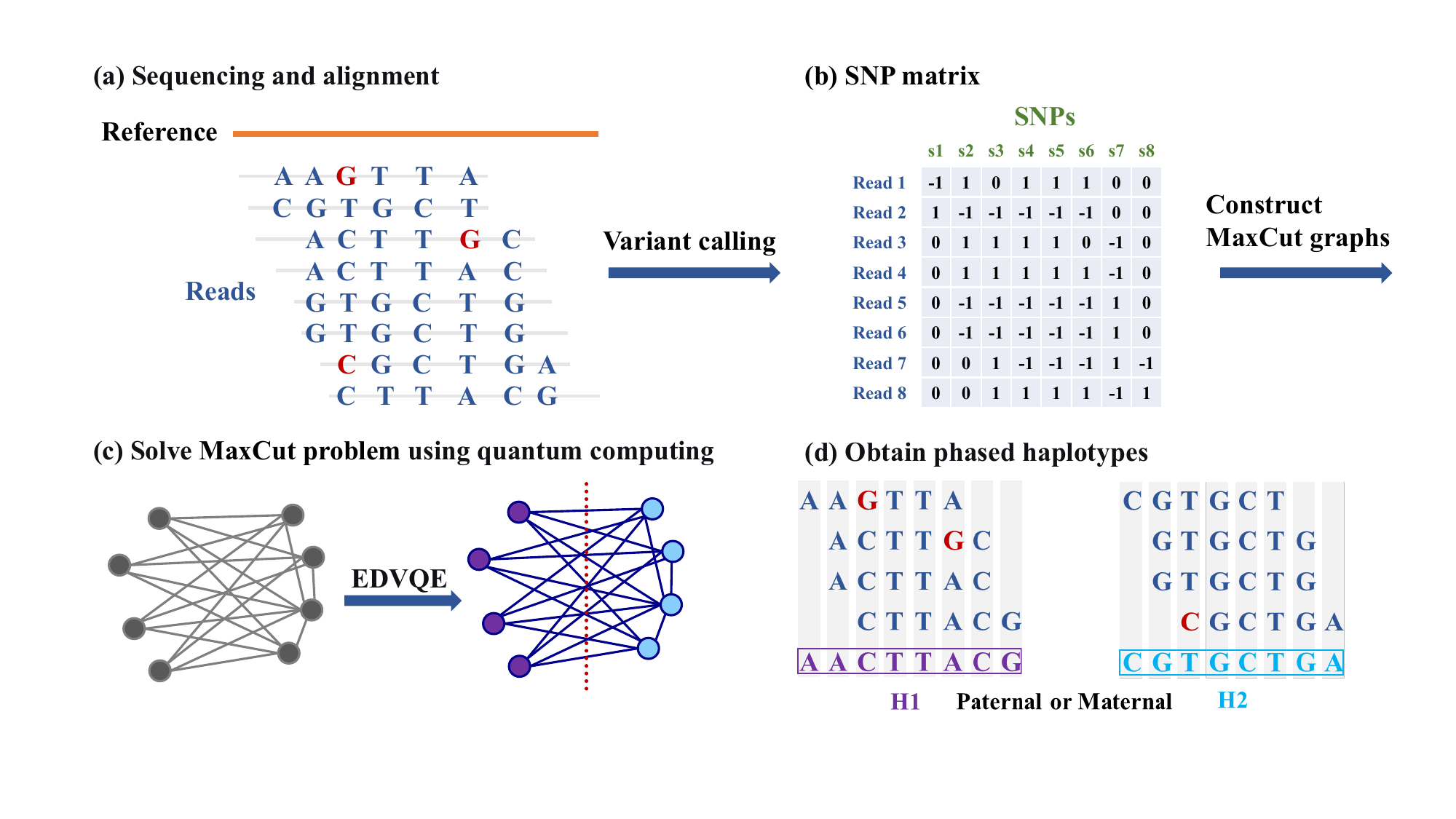}
\caption{Haplotype phasing flowchart. Panel (a) shows the alignment of sequenced reads to the reference genome. This is followed by (b) variant calling to generate an SNP matrix, where rows correspond to sequence fragments and columns to SNP loci. Matrix entries are encoded as 1 if the variant aligns with the allele in the candidate  haplotype H1 (CACTTAGG), -1 if it aligns with H2 (AGTGCTCA), and 0  if it aligns with neither or   the allele is inconclusive. Next, (c) MaxCut graphs are constructed from the SNP matrix, and the partitioning problem is solved using the EDVQE algorithm. Finally, (d) consensus analysis on the resulting two read partitions yields the phased haplotypes H1 and H2, assigned to paternal/maternal origins.}
\label{fig:Biological flowchart}
\end{figure}

We apply EDVQE to address haplotype phasing, a challenging combinatorial optimization problem in bioinformatics. Haplotype phasing can be reformulated as a MaxCut problem via graph-theoretic approaches. High-throughput sequencing fragments the homologous chromosomes into reads, which are represented as nodes in a weighted undirected graph. Edges connect reads that share overlapping heterozygous SNP sites, with edge weights determined by the number of discordant bases between the connected reads. We leverage our proposed EDVQE to maximize the cut value, which thereby effectively resolves conflicts between distinct haplotypes and facilitates the practical application of quantum algorithms. Subsequently, all reads are successfully classified into paternal or maternal groups based on SNP differences among the reads, and the final phased haplotype sequences are derived via consensus assembly. The workflow is illustrated in Figure \ref{fig:Biological flowchart}.

Experiments focus on the genomic region containing the \textit{ABCA1} gene, which spans 147,149 base pairs and contains 187 SNP sites. Located on chromosome 9q31.1, \textit{ABCA1} encodes a protein critical for reverse cholesterol transport and is implicated in diseases such as atherosclerosis and coronary artery disease (CAD) \cite{oram2005atp,haas2022potential,zaiou2018epigenetic,abdel2018increased,he2020diabetes,wu2022roles,fouladseresht2020association,yadav2023association,ren2022association,yang2022genetic}. All data are generated using CycloneSEQ sequencing technology from HG002 sample. For an \textit{ABCA1} haplotype phasing instance with 506 reads, we employ a distributed architecture consisting of 46 subsystems, each utilizing only 11 qubits. Table \ref{tab:performance} compares the performance of EDVQE with that of the GW algorithm across 50 trials with initial parameter seeds. EDVQE achieves a 40\% success rate in matching the cut value of the GW algorithm with $10^6$ projections, along with an average approximation ratio of 94.06\%.

\begin{table}[!ht]
\centering
\caption{Performance comparison between the EDVQE and the classical GW algorithm on a 506-node \textit{ABCA1} haplotype instance.}
\setlength{\tabcolsep}{12pt}
\begin{tabular}{lrrrrr}
\toprule
\multirow{2}{*}{\# of Nodes} & GW ($R = 10^6$) & \multicolumn{4}{c}{ EDVQE} \\ \cmidrule(lr){2-2} \cmidrule(lr){3-6}
 & Value & Best Value  & Average Value & Success Rate & Average Approximation Ratio \\
\midrule
506  & 103555 & 103555  & 97406.14 & 40.00\% & 94.06\% \\
\bottomrule
\end{tabular}
\label{tab:performance}
\end{table}

To assess phasing accuracy, the performance of EDVQE is  systematically compared against the gold-standard small variant benchmark dataset derived from the Genome in a Bottle Consortium. As shown in Table \ref{tab:Compared with the gold standard}, EDVQE achieves 100\% phasing completeness with zero switch error rate and Hamming error rate, with independent validation using the bioinformatics tool WhatsHap confirming identical results. The comparative analysis demonstrates that EDVQE is viable for real-world haplotyping applications, accurately phasing heterozygous SNP variants into fully resolved and error-free haplotypes comparable to established bioinformatics tools. It also holds significant potential for scaling to larger genomic datasets. This work validates EDVQE’s scalability and efficiency for genomic data while underscoring its utility in addressing complex genomics challenges, bridging quantum computing with practical bioinformatics workflows.

\begin{table}[!ht]
\centering
\caption{Performance comparison between the EDVQE and WhatsHap.}
\label{tab:Compared with the gold standard}
\renewcommand{\arraystretch}{1.3} 
\begin{tabular}{cccc}
\toprule
& Phasing Completeness &  Switch Error Rate & Hamming Error Rate \\
\midrule
EDVQE & 100\% & 0 & 0 \\
WhatsHap & 100\%  & 0 & 0 \\
\bottomrule
\end{tabular}
\end{table}

\label{sec1}

\section{Conclusion and Discussion}
In this work, we propose an enhanced distributed variational quantum eigensolver that integrates a distributed variational architecture with a novel hybrid classical-quantum perturbation strategy and warm-start initialization, thereby significantly extending the applicability of near-term quantum devices to large-scale MaxCut problems.

The core strength of EDVQE arises from the synergistic integration of its three principal components. First, the distributed strategy facilitates the resolution of large-scale problems under constrained qubit resources, enabling the successful tackling of MaxCut instances involving up to 1000 vertices with merely 10 qubits. Second, the hybrid classical-quantum perturbation method substantially enhances solution quality by merging targeted local search with expansive quantum-enabled exploration, thereby yielding outcomes that consistently outperform the classical GW algorithm, even when augmented by $10^4$ to $10^6$ random projections. In particular, the quantum two-vertex perturbation achieves markedly superior improvements in solution quality compared to classical single-vertex neighborhood search, with gains exceeding 25 times those of its classical counterpart on cluster graphs. Third, the warm-start initialization furnishes a high-quality and dependable initial configuration for the ensuing perturbation phase, collectively underpinning EDVQE's superior performance relative to the GW algorithm with $10^6$ projections.



The practical efficacy of this framework is further demonstrated through its successful application to the haplotype phasing problem, a computationally challenging task in genomics. Employing the EDVQE algorithm with only 11 qubits, we effectively addressed a 506-qubit haplotype phasing instance derived from real human \textit{ABCA1} gene sequencing data. The high-quality haplotype assemblies produced in this experiment underscore that the framework provides a feasible pathway for deploying hybrid classical-quantum solutions to large-scale problems in biological science and other engineering domains, even within the constraints of the NISQ era.


\section{Acknowledgements}

This work was supported by Shenzhen Science and Technology Program, China (JCYJ20241202123906009), Guangdong Provincial Quantum Science Strategic Initiative (Grant No. GDZX2503001).

\bibliography{sample}

\end{document}